\documentclass[runningheads]{llncs}

\usepackage[T1]{fontenc}
\usepackage{graphicx}
\usepackage{amsmath}
\usepackage{booktabs}
\usepackage{xcolor}
\usepackage{float}
\usepackage{url}
\title{Advancing LLM-based phoneme-to-grapheme for multilingual speech recognition}
\titlerunning{Advancing LLM-based P2G for Multilingual ASR}

\author{Lukuan Dong\inst{1} \and
Ziwei Li\inst{2} \and
Saierdaer Yusuyin\inst{4} \and
Xianyu Zhao\inst{1} \and
Zhijian Ou\inst{1,2,3}\thanks{Corresponding author (\email{ozj@tsinghua.edu.cn}). This work is supported by National Science and Technology Major Project (2023ZD0121401), Guangxi Science and Technology Project (GUIKEAB25069456).}}

\authorrunning{L. Dong et al.}

\institute{TasiTech Co., Ltd., China \and
{\scriptsize SPMI Lab, Tsinghua University, China\quad $^{3}$\enspace IEIT in Tianjin, Tsinghua University, China}%
\setcounter{@inst}{3}%
\and
School of Computer Science and Technology, Xinjiang University, China}

\newcommand{\revblue}[1]{{\color{black}#1}}
\newcommand{\ozja}{\textcolor{black}}
\newcommand{\ozjb}{\textcolor{black}}
\definecolor{myteal}{RGB}{0,128,128}
\newcommand{\dlk}{\textcolor{black}}

\begin{document}
\maketitle

% The abstract must exactly match the abstract entered into the paper submission system.
\begin{abstract}
Phoneme-based ASR factorizes recognition into speech-to-phoneme (S2P) and phoneme-to-grapheme (P2G), enabling cross-lingual acoustic sharing while keeping language-specific orthography in a separate module. While large language models (LLMs) are promising for P2G, multilingual P2G remains challenging due to language-aware generation and severe cross-language data imbalance. We study multilingual LLM-based P2G on the ten-language CV-Lang10 benchmark. 
\revblue{We examine DANP and Simplified SKM (S-SKM) as two robustness strategies for handling S2P uncertainty in the multilingual setting.} \ozjb{S-SKM is a Monte Carlo approximation that avoids CTC-based S2P probability weighting in P2G training.} 
\revblue{Robust training and low-resource oversampling reduce the average WER from 10.56\% to 7.61\%.}
\end{abstract}

\section{Introduction}

Phoneme-based automatic speech recognition (ASR) is a practical paradigm for multilingual and cross-lingual speech recognition \cite{li2020universal,zhu2021multilingual,xu2021simple,tjandra2023massively,lee2023optimizing,yusuyin2024whistle}. By modeling language-independent pronunciation units, such systems can share acoustic modeling capacity across languages and transfer knowledge from high-resource to low-resource settings. In this paradigm, recognition is typically decomposed into two stages: (i) Speech-to-Phoneme (S2P) acoustic modeling and (ii) Phoneme-to-Grapheme (P2G) conversion, where P2G maps phoneme sequences into written text.

P2G decoding is traditionally implemented using Weighted Finite State Transducers (WFSTs) that compose pronunciation lexicons with n-gram language models \cite{mohri2002wfst,chen1999smoothing,ou2010study}. While effective, WFST-based pipelines often require language-specific resources (lexicons, decoding graphs) and become cumbersome to scale to many languages or to rapidly evolving vocabularies. In multilingual deployments, using a language-specific WFST also typically assumes a prior language decision, while building a single unified decoding graph can become excessively large. Moreover, WFST-based approaches do not directly leverage the strong contextual and multilingual modeling capabilities of modern large language models (LLMs).

Motivated by this, recent work proposes LLM-based phoneme-to-grapheme (LLM-P2G) decoding by fine-tuning an LLM on paired phoneme and graphemic sequences \cite{ma2025llm}. To improve robustness to S2P errors and uncertainty, it further introduces Data Augmentation with Noisy Phonemes (DANP) and marginalization-based training methods such as Top-K Marginalization (TKM) \cite{ma2025llm} and later Sampling-K Marginalization (SKM) \cite{ma2025phoneme}. These methods have shown strong results in monolingual settings, and can potentially provide a graph-free alternative to WFST decoding.

\begin{figure}[t]
  \centering
  \includegraphics[width=1.0\textwidth]{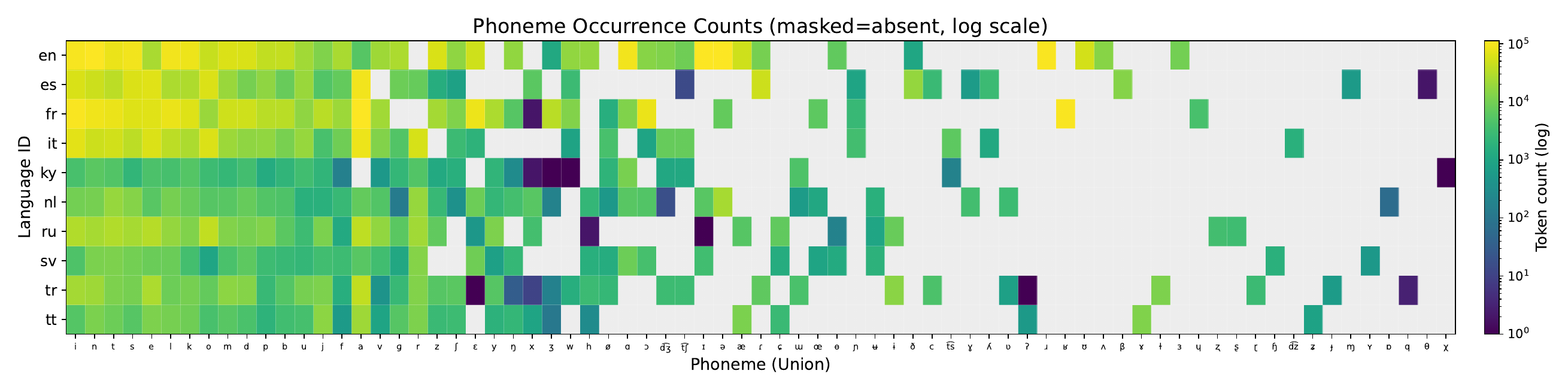}
  \caption{Phoneme occurrence counts across 10 languages in CV-Lang10. Each cell shows how many times a phoneme appears in the training data of a language.}
  \vspace{-6pt}
  \label{fig:phoneme_presence}
\end{figure}

However, extending LLM-based P2G to multilingual settings is challenging for at least two reasons. First, although phoneme sequences can be expressed in a universal symbol system (e.g., IPA) \cite{ipa1999handbook}, the model must generate language-specific orthographies, which requires language-aware generation and disambiguation. Second, multilingual corpora are often imbalanced: high-resource languages account for most training iterations, while low-resource languages contribute far fewer samples, which can lead to noticeable performance drops on those languages. Fig.~\ref{fig:phoneme_presence} further illustrates that many phonemes are shared across languages in CV-Lang10, while each language uses only a subset, highlighting the need to jointly model shared pronunciation patterns and language-specific spelling rules.

This paper advances LLM-based P2G for multilingual phoneme-based ASR. Using CV-Lang10, a ten-language Common Voice subset \cite{yusuyin2024whistle,ardila-etal-2020-common}, we build a single unified multilingual P2G model and systematically evaluate robustness and data-balancing strategies. 
\revblue{In the multilingual experiments, we position DANP and S-SKM as two robustness strategies for handling S2P uncertainty. S-SKM avoids calculating the CTC-based S2P probabilities used as weights in prior SKM-based P2G training, thereby slightly reducing training complexity.}
We further analyze low-resource oversampling as a mitigation for cross-language imbalance.

\revblue{Our contributions are: (1) we establish a unified multilingual LLM-P2G framework and benchmark it on CV-Lang10 under realistic cross-language imbalance; (2) we analyze DANP and S-SKM as two robustness strategies for multilingual P2G, both improving over direct multilingual fine-tuning in our setup; (3) we analyze the impact of cross-language data imbalance and show that a simple oversampling strategy improves low-resource languages.}

\section{Related Work}

\noindent\textbf{Phoneme-intermediate ASR.}
Two-stage ASR with phoneme intermediates has been studied as a way to improve multilingual transfer and modularity \cite{lee2023optimizing,yusuyin2024whistle}. In this setup, an acoustic model predicts phonemes (or phoneme lattices), and a second module maps them to written text. The separation allows the acoustic component to focus on pronunciation modeling while the text component captures orthographic and linguistic regularities.

\noindent\textbf{LLM-based P2G and robustness.}
LLMs have recently been used for post-processing and spelling correction in ASR pipelines \cite{li24h_interspeech,hu-etal-2024-listen}, and for direct phoneme-to-grapheme conversion \cite{ma2025llm}. A key practical issue is robustness to S2P errors and uncertainty. Ma et al. introduce DANP and Top-K Marginalization (TKM) \cite{ma2025llm}, which marginalize over the top-$K$ phoneme hypotheses from S2P beam search. A follow-up study proposes Sampling-K Marginalization (SKM) \cite{ma2025phoneme}, replacing top-$K$ enumeration with sampling from the CTC-based S2P distribution to better capture uncertainty. Marginalization-based objectives are attractive because they optimize the expected P2G likelihood under the S2P distribution, rather than training only on the 1-best hypothesis.

\noindent\textbf{Multilingual phoneme-to-grapheme conversion.}
Phoneme-to-grapheme conversion has been explored as a multilingual standalone task. Early work developed HMM-based multilingual P2G systems \cite{rentzepopoulos1996efficient}. More recently, PolyIPA trains a multilingual neural P2G model for transliteration and information retrieval across diverse writing systems \cite{lauc2024polyipa}. Compared with these conversion-focused studies, we consider an ASR setting where P2G must be robust to upstream S2P errors and trained under severe cross-language imbalance.

\section{Multilingual LLM-P2G}

\subsection{System Overview}

We use a decoupled two-stage architecture with an S2P acoustic model and an LLM-based P2G model. Given a speech feature sequence
$\mathbf{x} = (x_1, x_2,\dots, x_T)$, the system predicts a \emph{graphemic} sequence
$\mathbf{y} = (y_1, y_2,\dots, y_L)$ (formed by units such as characters or wordpieces) by marginalizing over intermediate phoneme
sequences $\mathbf{h}$:
\begin{equation}
p(\mathbf{y}|\mathbf{x}) = \sum_{\mathbf{h}} p(\mathbf{h}|\mathbf{x}) \, p(\mathbf{y}|\mathbf{h}),
\end{equation}
where $p(\mathbf{h}|\mathbf{x})$ is the S2P model and $p(\mathbf{y}|\mathbf{h})$ is the P2G model. In our experiments, the S2P stage uses Whistle \cite{yusuyin2024whistle} (Conformer encoder \cite{gulati20_interspeech} + CTC \cite{graves2006ctc}), and the P2G stage uses a decoder-only LLM (Qwen3 series \cite{qwen3technicalreport}) fine-tuned to generate language-specific text from phoneme inputs.

To support multilingual generation, we use a unified output space in which the P2G model predicts a language-identifier token \texttt{<lid>} followed by the graphemic sequence. This design encourages the model to jointly handle language identification (LID) and orthography generation.

\subsection{Robust P2G Training: DANP, TKM, SKM, and S-SKM}

Robust training aims to reduce the mismatch between noisy S2P outputs and the clean phoneme inputs typically used during P2G fine-tuning.

\noindent\textbf{Data Augmentation with Noisy Phonemes (DANP).}
DANP augments P2G fine-tuning with noisy phoneme inputs generated by the S2P model \cite{ma2025llm}. In practice, the noise can be constructed through several augmentation types, including N-best hypotheses from S2P beam search, random sampling from the S2P model, and aggregating hypotheses produced by different S2P checkpoints to increase diversity. DANP is simple and effective, but it may require manual tuning of augmentation types and their hyper-parameters (e.g., beam width, sampling number, and the number of checkpoints), which can be non-trivial in multilingual settings.

\noindent\textbf{Top-K Marginalization (TKM).}
Instead of training on a single best phoneme hypothesis, TKM considers a set of top-$K$ candidate phoneme sequences \cite{ma2025llm}
$\mathcal{H}= \{\mathbf{h}^{(1)},\dots,\mathbf{h}^{(K)}\}$ (typically obtained by S2P beam search)  and approximates the marginal likelihood by
\begin{equation}
\begin{aligned}
p(\mathbf{y}|\mathbf{x})
&\approx \sum_{\mathbf{h}\in \mathcal{H}} p(\mathbf{h}|\mathbf{x})\, p(\mathbf{y}|\mathbf{h}) \\
&= \sum_{k=1}^{K} p(\mathbf{h}^{(k)}|\mathbf{x}) \prod_{i=1}^{L} p(y_i \,|\, \mathbf{h}^{(k)}, \mathbf{y}_{1:i-1}),
\end{aligned}
\label{eq:tkm}
\end{equation}
where $p(\mathbf{h}^{(k)}|\mathbf{x})$ is the S2P probability of the $k$-th hypothesis, and the P2G model provides autoregressive probabilities over graphemic units.

\noindent\textbf{Sampling-K Marginalization (SKM).}
SKM replaces beam search with stochastic sampling from the CTC-based S2P model to generate diverse phoneme hypotheses \cite{ma2025phoneme}. Specifically, it draws a frame-level CTC path $\boldsymbol{\pi}^{(k)}$ by sampling one symbol (including blank) at each frame from the CTC distribution, and then converts the path into a phoneme sequence $\mathbf{h}^{(k)} = \mathcal{B}(\boldsymbol{\pi}^{(k)})$ using the standard CTC mapping function $\mathcal{B}(\cdot)$. Here, $\mathcal{B}(\cdot)$ first collapses consecutive repeated non-blank symbols and then removes blank symbols. Compared with TKM, where $\{\mathbf{h}^{(k)}\}$ are the top-$K$ hypotheses from beam search, SKM yields samples that better cover the S2P uncertainty region and obtains better performance \cite{ma2025phoneme}. 
However, \ozja{as can be seen from Eq.~(\ref{eq:tkm})}, SKM still requires computing the CTC-based S2P probability $p(\mathbf{h}^{(k)}|\mathbf{x})$ for each sampled hypothesis (via the CTC forward algorithm) to use as a weight, which \ozja{adds overhead} in large multilingual corpora.

\noindent\textbf{Simplified Sampling-K Marginalization (S-SKM).}
\revblue{We examine Simplified SKM (S-SKM) as a streamlined SKM variant for large-scale multilingual settings.} Instead of calculating the CTC-based S2P probabilities used as per-hypothesis weights in SKM, S-SKM approximates the path marginalization via Monte Carlo sampling:
\begin{equation}
\begin{aligned}
p(\mathbf{y}|\mathbf{x})
&= \sum_{\boldsymbol{\pi}} p(\boldsymbol{\pi}|\mathbf{x}) \, p(\mathbf{y}|\mathcal{B}(\boldsymbol{\pi})) \\
&\approx \frac{1}{K}\sum_{k=1}^{K} p\left(\mathbf{y}\,|\,\mathcal{B}(\boldsymbol{\pi}^{(k)})\right),
\end{aligned}
\label{eq:sskm}
\end{equation}
where $\boldsymbol{\pi}^{(k)} \sim p(\boldsymbol{\pi}|\mathbf{x})$ is a sampled CTC state path (same length as $\mathbf{x}$) and $\mathcal{B}(\cdot)$ maps a CTC state path to a phoneme sequence. Eq.~(\ref{eq:sskm}) is an unbiased Monte Carlo estimator of the marginal likelihood under the CTC path distribution, and removes the need for running the forward algorithm to calculate the S2P probability during P2G training. In \S\ref{sec:results} we show that S-SKM closely matches SKM in monolingual ablations, while being simpler in multilingual settings.

\subsection{P2G Decoding}
For decoding we follow a TKM-style rescoring scheme \cite{ma2025llm}. \revblue{This decoding procedure is applied consistently to all P2G models, regardless of whether they are trained with DANP, SKM, or S-SKM; these methods differ only in the robustness training strategy.} We run S2P beam search to obtain top-$K$ phoneme hypotheses $\{\mathbf{h}^{(k)}\}_{k=1}^{K}$. For each
$\mathbf{h}^{(k)}$, the P2G model generates top-$S$ text candidates, yielding up to $K \times S$ candidates that are deduplicated into a set $\mathbf{Y}$. Each
$\mathbf{y}\in \mathbf{Y}$ is then scored using Eq.~(\ref{eq:tkm}), using beam scores as $p(\mathbf{h}^{(k)}|\mathbf{x})$ and setting $p(\mathbf{y}|\mathbf{h}^{(k)}) \approx 0$ if $\mathbf{y}$ is not among the $S$ candidates generated from $\mathbf{h}^{(k)}$.

\subsection{Multilingual Adaptation and Data Balancing}

% Multilingual training is affected by severe cross-language imbalance. Let $H_l$ denote the number of training hours for language $l$. We use a simple oversampling strategy that enforces a minimum target exposure per language:
% \begin{equation}
% \tilde{H}_l = \max(H_l, H_{\text{tgt}}).
% \label{eq:oversample}
% \end{equation}
% \ozja{We empirically set $H_{\text{tgt}} = 240~\text{hours}$, since each of the four high-resource languages has at least 271.5 hours of training data (Italian is the smallest at 271.5 hours), and 240 hours provides a conservative balancing target.
% For the six low-resource languages (ky, nl, ru, sv, tr, tt), where $H_l < H_{\text{tgt}}$, we repeatedly sample training utterances until their effective hours reach $H_{\text{tgt}}$.
% High-resource languages are left unchanged. This increases the update frequency of low-resource languages during optimization without changing the model architecture, and is complementary to robustness training.}

Multilingual training is affected by severe cross-language imbalance. Let $H_l$ denote the number of training hours for language $l$. We use a simple oversampling strategy that \dlk{repeats each language by an integer factor relative to Italian}:
\begin{equation}
\tilde{H}_l = \dlk{\left\lceil \frac{H_{\mathrm{it}}}{H_l} \right\rceil H_l}.
\label{eq:oversample}
\end{equation}
\ozja{We empirically \dlk{use Italian as the reference language}, since as shown in Table \ref{tab:multilingual_results}, each of the four high-resource languages has at least 271.5 hours of training data (Italian is the smallest with $H_{\mathrm{it}}=$271.5 hours)\dlk{. For each language $l$, we set its repetition factor to $\left\lceil H_{\mathrm{it}} / H_l \right\rceil$, i.e., the smallest integer multiple needed to reach at least the Italian level.}
For the six low-resource languages (ky, nl, ru, sv, tr, tt), \dlk{this means repeatedly sampling training utterances by the corresponding integer multiple.}
High-resource languages are left unchanged. This increases the update frequency of low-resource languages during optimization without changing the model architecture, and is complementary to robustness training.}

\section{Experimental Setup}

\subsection{Datasets}

We evaluate on CV-Lang10 (English-en, Spanish-es, French-fr, Italian-it, Kyrgyz-ky, Swedish-sv, Dutch-nl, Russian-ru, Turkish-tr, Tatar-tt), a ten-language subset of Common Voice with additional data cleaning and pronunciation lexicon construction \cite{yusuyin2024whistle,ardila-etal-2020-common}. The per-language training hours are shown in Table~\ref{tab:multilingual_results}. We report S2P phoneme error rate (PER) and final ASR word error rate (WER) after P2G. For monolingual ablations, we additionally use the Common Voice v11.0 Polish (pl) and German (de) subsets following \cite{ma2025llm}.

\subsection{Models}

\noindent\textbf{S2P model.}
In multilingual experiments, we use Whistle-large (543M parameters) as the S2P model for both training and evaluation \cite{yusuyin2024whistle}. It is a multilingual phoneme recognizer with a Conformer encoder and a CTC objective.

\noindent\textbf{P2G model.}
We use decoder-only Qwen3 models (Qwen3-1.7B-Base and Qwen3-4B-Base) as the P2G backbones \cite{qwen3technicalreport}. 
% The training serialization (including the language identifier token) is described in \S\ref{sec:training}.

\subsection{Training Details}\label{sec:training}

All P2G models are optimized using Adam \cite{kingma2015adam} with a warmup + cosine decay schedule. The learning rate is linearly increased from $1\times10^{-4}$ to a peak value of $2\times10^{-4}$ during the first 10\% of training steps, and then annealed with cosine decay. Training is conducted on 4$\times$ NVIDIA A800 GPUs with a micro-batch size of 8 per GPU, resulting in a global batch size of 32; we then accumulate gradients for 16 steps to obtain an effective batch size of 512.

Training samples are serialized as:
\begin{center}
\vspace{-0.2cm}
\small
\textbf{\texttt{<ipa> \{p\} | <lid> \{g\}}}
\vspace{-0.2cm}
\end{center}
where $\{p\}$ is the input phoneme sequence and the model autoregressively generates the language identifier \texttt{<lid>} followed by the graphemic sequence $\{g\}$.

We first conduct monolingual ablations on Polish (pl) and German (de) to compare SKM and S-SKM under controlled settings (both with $K=8$). For this study, models are trained for 20 epochs to ensure reliable comparison. We then evaluate multilingual training on CV-Lang10.

For DANP, we generate 16-best phoneme hypotheses per utterance using CTC beam search on the S2P model and treat them as noisy inputs for P2G fine-tuning \cite{ma2025llm}. For S-SKM, we obtain $K=8$ samples per utterance by drawing state paths from the S2P CTC distribution (including blanks), and then applying the CTC mapping function $\mathcal{B}(\cdot)$. \dlk{Specifically, following the
SKM temperature-sampling strategy~\cite{ma2025phoneme}, we divide the CTC
logits by a temperature $T$ before softmax sampling and set $T=1.5$ for
both SKM and S-SKM training.
}

\ozjb{To ensure a fair comparison across training strategies, we control the overall optimization budget: although DANP and S-SKM construct training examples differently, we match the total number of effective updates. In particular, DANP is trained for 2 epochs and S-SKM for 4 epochs, keeping the numbers of training instances and optimization steps comparable.}

For LLM fine-tuning, we use LoRA-based parameter-efficient adaptation \cite{hu2022lora}.
\ozja{All experiments use the same LoRA configuration: rank $r=256$ is applied to $q\_proj$, $k\_proj$, $v\_proj$, $o\_proj$, $gate\_proj$, $down\_proj$, and $up\_proj$ in every Transformer block, resulting in 528M trainable parameters.}

% Across experiments, we keep the trainable adapter parameter count fixed (about 0.5B parameters) to ensure comparable adaptation capacity across backbones and training strategies.

\section{Experimental Results}
\label{sec:results}

We first report the monolingual ablation between SKM and S-SKM, and then evaluate multilingual P2G under different robustness and data-balancing strategies.

\begin{table}[t]
  \caption{
  Ablation study of training strategies (SKM vs. S-SKM).
  The P2G model is fixed to Qwen3-1.7B-Base. Results are reported in WER (\%).
  We also report the S2P PER (\%) on the same test sets.
  }
  \label{tab:skm_ablation}
  \centering
  \small
  \renewcommand{\arraystretch}{1.05}
  \begin{tabular}{l cc}
    \toprule
    \textbf{Training Method} & \textbf{pl-test} & \textbf{de-test} \\
    \midrule
    SKM WER   & 4.25 & \textbf{11.98} \\
    S-SKM WER & \textbf{4.14} & 12.13 \\
    \midrule
    \multicolumn{1}{l}{\textbf{S2P PER (\%)}} & 1.97 & 5.37 \\
    \bottomrule
  \end{tabular}
\end{table}

\subsection{Monolingual Ablation}

Table~\ref{tab:skm_ablation} compares SKM and S-SKM on Polish and German with Qwen3-1.7B-Base as the P2G backbone. Following prior work \cite{ma2025llm}, we fine-tune the Whistle-small CTC model to obtain the S2P component for these two languages, and apply TKM decoding for evaluation.

The two strategies yield nearly identical performance. On \textit{pl-test}, S-SKM achieves 4.14\% WER, slightly better than SKM (4.25\%), while on \textit{de-test}, SKM attains 11.98\% WER, marginally better than S-SKM (12.13\%). The absolute gap is within 0.15\% WER, indicating that equal-weight Monte Carlo marginalization is a reasonable approximation in practice. From a systems perspective, in our setup (batch size = 512), computing CTC probabilities for sampled hypotheses (8 per utterance) takes about 8 ms per iteration on 4$\times$ NVIDIA A800 GPUs. \ozja{Across 33K iterations, a typical setting of a single run of our multilingual experiments, this introduces a small overhead.} 
\revblue{Given the comparable accuracy and better efficiency, we include S-SKM as a simplified robustness strategy in the multilingual experiments.}

\subsection{Multilingual P2G Performance}
\vspace{-3pt}

\begin{table}[t]
\caption{
Multilingual results across 10 languages (CV-Lang10).
Training hours per language are shown in the first row.
PER (\%) of Whistle-large (543M) is reported,
followed by P2G WER (\%) results. All P2G models use top-$K$ marginalization decoding with $K{=}8$.
The best WER result in each column is \textbf{boldfaced},
and the second-best WER is \underline{underlined}.
The last column reports the training-hours-weighted average WER.
}
\vspace{-8pt}
\label{tab:multilingual_results}
\centering
\scriptsize
\renewcommand{\arraystretch}{1.10}
\setlength{\tabcolsep}{4pt}

\resizebox{\textwidth}{!}{%
\begin{tabular}{l c c c c c c c c c c c c}
\toprule
Model
& en & es & fr & it & ky & nl & ru & sv & tr & tt & avg & hrs-wavg \\
\midrule

\textit{Training Hours}
& 2227.3 & 382.3 & 823.4 & 271.5 & 32.7 & 70.2 & 149.8 & 29.8 & 61.5 & 20.8 & -- & -- \\

\midrule
\multicolumn{13}{l}{\textbf{S2P model (PER\%)}}\\
\midrule

Whistle-large
& 5.42 & 1.96 & 3.52 & 2.25 & 4.06 & 2.64 & 2.97 & 11.33 & 4.04 & 5.97 & 4.41 & 4.37 \\

\midrule
\multicolumn{13}{l}{\textbf{P2G model (WER\%)}}\\
\midrule

(E1) Qwen3-4B-Base
& 8.26 & \textbf{5.84} & 10.44 & 6.84
& 10.07 & 6.05 & 5.98 & 17.94 & 10.92 & 23.25 & 10.56 & 8.46 \\

(E2) \hspace{1em}+ DANP
& \underline{8.04} & 6.29 & \underline{10.31} & \underline{6.74}
& 5.14 & \underline{5.33} & 3.91 & 14.56 & 7.84 & 16.41 & 8.45 & \underline{8.11} \\

\dlk{(E3) \hspace{1em}+ DANP + oversampling}
& \textbf{7.95} & \underline{5.96} & \textbf{10.13} & \textbf{6.65}
& 3.97 & \textbf{4.96} & \underline{3.85} & 12.40 & \textbf{6.99} & \underline{13.25} & \underline{7.61} & \textbf{7.93} \\

(E4) \hspace{1em}+ S-SKM
& 8.24 & 6.15 & 10.53 & 7.18
& 7.07 & 6.03 & 4.95 & 15.88 & 9.78 & 20.65 & 9.64 & 8.41 \\

(E5) \hspace{1em}+ S-SKM + oversampling
& 8.22 & 6.12 & 10.62 & 7.17
& \underline{2.82} & 5.48 & 3.90 & \underline{10.59}
& \underline{7.33} & 14.34 & 7.66 & 8.22 \\

\midrule
(E6) WFST baseline
& 8.80 & 7.02 & 14.02 & 8.16
& \textbf{0.94} & 6.22 & \textbf{1.46} & \textbf{5.06}
& \underline{7.05} & \textbf{6.92} & \textbf{6.56} & 9.20 \\

\bottomrule
\end{tabular}%
}
\end{table}

{
\setlength{\textfloatsep}{6pt plus 2pt minus 2pt}
\begin{table}[t]
\caption{
LID accuracy (\%) for multilingual P2G models (E1--E5).
The last two columns report the macro average accuracy and the training-hours-weighted
average accuracy, weighted by the per-language training hours. (\textbf{bold}: best)
}
\vspace{-10pt}
\label{tab:multilingual_lid_acc}
\centering
\scriptsize
\renewcommand{\arraystretch}{1.10}
\setlength{\tabcolsep}{4pt}
\resizebox{\textwidth}{!}{%
\begin{tabular}{l c c c c c c c c c c c c}
\toprule
Model
& en & es & fr & it & ky & nl & ru & sv & tr & tt & avg & hrs-wavg \\

\midrule
(E1) Qwen3-4B-Base
& 99.72 & 99.54 & \textbf{99.80} & 99.05 & 99.01 & 99.73 & 99.44 & 97.91 & 98.88 & 95.63 & 98.87 & 99.61 \\
(E2) \hspace{1em}+ DANP
& 99.69 & 99.54 & 99.78 & \textbf{99.11} & \textbf{99.13} & 99.71 & 99.46 & 97.67 & 98.84 & 96.25 & 98.92 & 99.60 \\
\dlk{(E3) \hspace{1em}+ DANP + oversampling}
& 99.71 & 99.55 & \textbf{99.80} & 98.85 & 99.01 & \textbf{99.76} & 99.40 & \textbf{98.22} & \textbf{99.07} & \textbf{96.64} & \textbf{99.00} & 99.60 \\
(E4) \hspace{1em}+ S-SKM
& \textbf{99.73} & \textbf{99.59} & 99.78 & 99.09 & 98.76 & 99.69 & 99.54 & 97.95 & 98.97 & 95.77 & 98.89 & \textbf{99.62} \\
(E5) \hspace{1em}+ S-SKM + oversampling
& 99.68 & 99.57 & 99.78 & 99.07 & 98.70 & 99.69 & \textbf{99.59} & 97.69 & 99.05 & 95.18 & 98.80 & 99.59 \\
\bottomrule
\end{tabular}%
}
\vspace{-5pt}
\end{table}
}

Table~\ref{tab:multilingual_results} reports multilingual results on CV-Lang10. Table~\ref{tab:multilingual_lid_acc} \dlk{shows that all multilingual settings maintain high LID accuracy, with macro averages between 98.80\% and 99.00\%,} indicating that the unified model can reliably infer language from phoneme input. \dlk{Even when oversampling slightly lowers macro LID in the S-SKM setting (E5 vs.\ E4), WER still improves, suggesting that the remaining errors are dominated by orthographic generation rather than language identification.}

\noindent\textbf{Direct multilingual fine-tuning.}
E1 extends monolingual LLM-based P2G to multilingual training without explicit robustness modeling. It performs reasonably on high-resource languages (en, es, fr, it), but degrades on low-resource languages such as Kyrgyz (10.07\%) and Tatar (23.25\%). This suggests that naive multilingual fine-tuning is sensitive to cross-language imbalance and can under-train low-resource orthographic patterns.

\noindent\textbf{DANP and S-SKM.}
\ozja{\dlk{E2 and E4 evaluate DANP and S-SKM as two robustness strategies in the multilingual setting. Both improve over E1, suggesting that modeling S2P uncertainty is useful beyond the monolingual case. 
Without oversampling, DANP is stronger in this setup, reducing the average WER to 8.45\% compared with 9.64\% for S-SKM. Their detailed WER patterns vary across languages, and Table~\ref{tab:multilingual_results} reports the full results.}}
We also observe that languages with higher S2P PER (e.g., Swedish) tend to have higher downstream WER, and robustness training helps but does not fully remove this dependency, indicating that improving S2P remains complementary.

\noindent\textbf{Effect of low-resource oversampling.}
\dlk{To mitigate imbalance during multilingual training, we further apply the oversampling strategy in Eq.~(\ref{eq:oversample}) on top of both robustness methods. For DANP, E3 reduces the unweighted average WER from 8.45\% to 7.61\% and the training-hours-weighted average WER from 8.11\% to 7.93\%, with gains on low-resource languages such as Kyrgyz (5.14\% $\rightarrow$ 3.97\%) and Tatar (16.41\% $\rightarrow$ 13.25\%). For S-SKM, E5 reduces the unweighted average WER from 9.64\% to 7.66\%, with clear gains on low-resource languages (ky: 7.07\% $\rightarrow$ 2.82\%, sv: 15.88\% $\rightarrow$ 10.59\%, tt: 20.65\% $\rightarrow$ 14.34\%). With oversampling, DANP slightly outperforms S-SKM in the unweighted average WER (7.61\% vs.\ 7.66\%) and the training-hours-weighted average WER (7.93\% vs.\ 8.22\%). Oversampling also does not harm high-resource languages: en/es/fr/it remain close to their corresponding non-oversampled systems, suggesting that increased low-resource exposure mainly reduces bias and interference rather than overfitting.}

\noindent\textbf{Comparison with WFST.}
\dlk{E6, the WFST baseline, achieves the best unweighted average WER (6.56\%), with good results on low-resource languages such as Kyrgyz and Tatar, consistent with the benefit of language-specific lexicons and decoding graphs. In contrast, our multilingual LLM-P2G uses a single unified P2G model and does not depend on per-language WFST infrastructure. The E3 LLM-P2G model matches or exceeds the WFST baseline on several languages (en, es, fr, it, nl, tr) and achieves a better training-hours-weighted average WER (7.93\% vs.\ 9.20\%). This highlights a trade-off: WFST remains stronger on the lowest-resource languages, while unified LLM-based P2G provides a graph-free solution with competitive performance where training data is sufficient.}

% \subsubsection{Discussion}

% The results also reveal several practical considerations for deploying multilingual LLM-P2G. First, the unified model is highly competitive on high-resource languages and improves the training-hours-weighted average WER, but the lowest-resource languages still benefit strongly from language-specific constraints (WFST) and may require additional supervision beyond simple repetition. A promising direction is to incorporate lightweight lexical constraints or external text resources for these languages while keeping the overall decoding pipeline graph-free and easier to maintain.

% Second, training efficiency matters in the multilingual setting. Compared with SKM, S-SKM avoids per-hypothesis weighting through the CTC forward algorithm and therefore reduces the computational overhead when scaling to more languages and larger datasets. Together with oversampling, this enables robust multilingual training without introducing language-specific components.

% Finally, we keep the S2P model fixed in this work. For languages with high S2P PER, downstream WER remains relatively high even after robust P2G training, suggesting that future gains may come from jointly improving S2P and P2G, for example via targeted S2P adaptation or end-to-end fine-tuning with a P2G-aware objective.

\section{Conclusion and Future Work}
\vspace{-3pt}

We presented a unified multilingual LLM-based phoneme-to-grapheme module for phoneme-based ASR and a systematic study on CV-Lang10. 
\revblue{We studied DANP and S-SKM as two robustness strategies in the multilingual setting, with S-SKM serving as a simplified Monte Carlo approximation to SKM that removes CTC probability weighting.} \dlk{Experiments show that naive multilingual fine-tuning is vulnerable to cross-language imbalance. 
Robustness training with DANP and S-SKM strategies and simple low-resource oversampling yield improvements, with DANP + oversampling giving the best unified P2G result and reducing average WER from 10.56\% to 7.61\%.}
% \ozjb{DANP and S-SKM can both be viewed as robustness strategies for multilingual P2G under S2P uncertainty.}

One takeaway is that the bottleneck in multilingual phoneme-based ASR depends on the language. For some low-resource languages with moderate S2P PER (e.g., ky), the main limitation is P2G data scarcity and imbalance, and oversampling is effective. For languages with high S2P PER (e.g., sv), robustness and oversampling improve performance but S2P errors remain a limiting factor, suggesting that S2P quality and P2G robustness should be considered jointly in future system design.
In this regard, incorporating pronunciation-lexicon free end-to-end training of both S2P and P2G models \cite{jsaspg} is potentially interesting future work to further improve LLM-based multilingual speech recognition.

% Future work includes: (i) improving language control (e.g., conditioning on known language ID or lightweight language adapters) to reduce cross-language interference, (ii) incorporating external text or lexicon resources to better handle rare words and spelling conventions in low-resource languages, (iii) studying more principled imbalance-aware optimization beyond simple repetition, and (iv) improving decoding efficiency by reducing the candidate generation cost in the $K \times S$ rescoring scheme without sacrificing accuracy.

\newpage
\section{Generative AI Use Disclosure}
Generative AI tools are used in this work only for language editing, polishing, and formatting of the manuscript. They are not
used to generate any core content, research ideas, experimental
designs, results, or major textual parts of the paper. All scientific contributions, including model design, experiments, analysis, and conclusions, are completed by the authors.

\bibliographystyle{splncs04}
\bibliography{mybib_formatted}

\end{document}